\documentclass[twocolumn,showpacs,aps,prl,superscriptaddress]{revtex4}

\usepackage{graphicx}
\usepackage{dcolumn}
\usepackage{amsmath}
\usepackage{epsfig}

\input pubboard/babarsym
\newcommand{\bi}{\begin{itemize}}
\newcommand{\ei}{\end{itemize}}
\newcommand{\ben}{\begin{enumerate}}
\newcommand{\een}{\end{enumerate}}
\newcommand{\bc}{\begin{center}}
\newcommand{\ec}{\end{center}}
\newcommand{\bt}{\begin{table}}
\newcommand{\et}{\end{table}}
\newcommand{\be}{\begin{equation}}
\newcommand{\eeq}{\end{equation}}
\newcommand{\ba}{\begin{eqnarray}}
\newcommand{\ea}{\end{eqnarray}}
\newcommand{\hs}{\hspace}

\newcommand{\la}{\ifmmode {\leftarrow} \else {$\leftarrow$}\fi}
\newcommand{\Ra}{\ifmmode {\Rightarrow} \else {$\Rightarrow$}\fi}
\newcommand{\La}{\ifmmode {\Leftarrow} \else {$\Leftarrow$}\fi}
\newcommand{\Lra}{\ifmmode {\Longrightarrow} \else {$\Longrightarrow$}\fi}
\newcommand{\Lla}{\ifmmode {\Longleftarrow} \else {$\Longleftarrow$\fi}}
\newcommand{\Llra}{\ifmmode {\Longleftrightarrow} \else {$\Longleftrightarrow$\fi}}
\newcommand{\Lk}{\ifmmode {{\cal L}} \else {${\cal L}$}\fi}
\newcommand{\Wt}{\ifmmode {{\cal W}} \else {${\cal W}$}\fi}
\newcommand{\Br}{\ifmmode {{\cal B}} \else {${\cal B}$}\fi}
\newcommand{\N}{\ifmmode {{\cal N}} \else {${\cal N}$}\fi}
\newcommand{\G}{\ifmmode {{\cal G}} \else {${\cal G}$}\fi}
\newcommand{\E}{\ifmmode {{\cal E}} \else {${\cal E}$}\fi}

\newcommand {\glt} {\hspace{1mm}\raisebox{-0.10cm}[1.5ex][0in]{$\sim$}
  \hspace{-3.3mm}\raisebox{0.08cm}[1.5ex][0in]{$>$}\hspace{1mm}}
\newcommand{\tBz}{\ifmmode {\tau_{\Bz}} \else {$\tau_{\Bz}$ }\fi }
\newcommand{\tBp}{\ifmmode {\tau_{\Bu}} \else {$\tau_{\Bu}$ }\fi }

\newcommand{\BtoDm}{\mbox{$B^0 \rightarrow D^{*-} \ell^+ \nu_\ell$}}

\newcommand{\pstar}{\ifmmode {\pi_s} \else {$\pi_s$}\fi }
\newcommand{\plab}{\ifmmode{p} \else {$p$}\fi}
\newcommand{\ks}{\ifmmode{k^*} \else {$k^*$}\fi}
\newcommand{\mnusq}{\ifmmode{{M_\nu}^2} \else {${M_{\nu}}^2$}\fi} 
\newcommand{\DTau}{\ifmmode {\Delta \tau} \else {$\Delta \tau$}\fi}
\newcommand{\ggcc}{\ifmmode {GeV^2/c^4} \else {$GeV^2/c^4$}\fi}
\def\BpBm {\ensuremath{B^+ {\kern -0.16em \Bub}}}

\newcommand{\BABARPubYear}    {02}
\newcommand{\BABARPubNumber}  {02}

\newcommand{\SLACPubNumber} {9128}

\def\figurebox#1#2#3{%
    \def\arg{#3}%
    \ifx\arg\empty
    {\hfill\vbox{\hsize#2\hrule\hbox to #2{\vrule\hfill\vbox to #1{\hsize#2\vfill}\vrule}\hrule}\hfill}%
    \else
    {\hfill\epsfbox{#3}\hfill}%
    \fi}

\begin{document}

\preprint{\babar-PUB-\BABARPubYear/\BABARPubNumber} 
\preprint{SLAC-PUB-\SLACPubNumber} 

\begin{flushleft}
\babar-PUB-\BABARPubYear/\BABARPubNumber\\
SLAC-PUB-\SLACPubNumber\\
\end{flushleft}

\title{
{\large \bf Measurement of the {\boldmath \Bz} Lifetime 
with Partially Reconstructed 
\mbox{{\boldmath $\Bz\ \ra \dsm \ell^+ \nu_{\ell}$}} Decays}
}

%
\author{B.~Aubert}
\author{D.~Boutigny}
\author{J.-M.~Gaillard}
\author{A.~Hicheur}
\author{Y.~Karyotakis}
\author{J.~P.~Lees}
\author{P.~Robbe}
\author{V.~Tisserand}
\author{A.~Zghiche}
\affiliation{Laboratoire de Physique des Particules, F-74941 Annecy-le-Vieux, France }
\author{A.~Palano}
\author{A.~Pompili}
\affiliation{Universit\`a di Bari, Dipartimento di Fisica and INFN, I-70126 Bari, Italy }
% ...
\collaboration{The \babar\ Collaboration}
\noaffiliation

\date{\today}%

\begin{abstract}
The \Bz\ lifetime has been measured with a sample of 23 million \BB\ pairs
collected by the \babar\ detector at the \pep2\ $e^+ e^-$ storage ring
during 1999 and 2000. Events from the semileptonic decay \BtoDm\
have been selected with a partial reconstruction 
method in which only the charged lepton and the slow $\pi$ from the \mbox{$\dsm \ra \Dzb \pi^-$} decay are reconstructed. The result is
$$\tBz = 1.529 \pm 0.012 ~\mathrm{(stat)}\pm 0.029 ~\mathrm{(syst)} ~\ps.$$ 
\end{abstract}


\maketitle

The technique of partial reconstruction of  \dsm mesons (charge conjugate states are always implied), 
in which only the slow pion from the \mbox{$\dsm \ra \Dzb \pi^-$} decay is required,
has been widely used in the past \cite{INCL} to select large samples of reconstructed $B$ mesons. 
This technique provides 
a way to measure the combination of CKM angles $(2\beta +\gamma)$ with
\mbox{$\Bz \ra \dsm \pi^+$} decays \cite{PhBook}. 
The application of this method to the semileptonic decay
\mbox{\BtoDm} allows the method's validation, while providing a tool
for precise measurements of several properties of the \Bz , including
its lifetime, \tBz , and the \BzBzb mixing parameter, $\Delta m_d$.
A precise measurement of \tBz\ is presented herein. \par
The data used in this analysis, recorded by the \babar\ detector
at the \pep2\ storage ring during  1999-2000, correspond to an integrated luminosity of 
20.7 \invfb\ collected on the \FourS\ resonance (on-peak events) and 2.6 \invfb\ collected 40 MeV 
below the resonance (off-peak) for background studies. 
Samples of simulated \BB\ events were analyzed through the same
analysis chain as the real data. The equivalent luminosity of the simulated data is approximately 
equal to the on-peak data. \par
A detailed description of the  \babar\ detector and the algorithms used for track reconstruction, particle 
identification, and selection of \BB\ events is provided elsewhere~\cite{ref:babar}; a brief
summary is given here.
Particles with momenta $\plab \glt 170 \mevc$
are reconstructed by matching hits in the silicon vertex tracker (SVT) 
with track elements in the drift chamber (DCH). Since lower momentum tracks do not 
leave signals on many wires in the DCH due to the bending induced by the magnetic field, they
 are  reconstructed by the SVT alone. 
Electrons are identified with the ratio of the track momentum to the associated energy deposited in the calorimeter (EMC), 
the transverse profile of the shower, the energy loss in the drift chamber, and the information
from the Cherenkov detector (DIRC). The efficiency for electron identification in the acceptance of 
the electromagnetic calorimeter is about $90\%$, 
with a hadron misidentification probability equal to  $0.15$\%.
Muon candidates are required to have a path length and hit distribution in the instrumented 
flux return and energy deposition in the EMC consistent with that expected for a minimum-ionizing particle.
The Cherenkov light emission in the DIRC 
is then employed to further reject kaons misidentified as muons, by requiring muon candidates
to have a kaon hypothesis probability less than 5\%.
These criteria yield $74\%$ muon efficiency with $2.6\%$ hadron misidentification probability. \par
Semileptonic \Bz\ decays are then selected by searching for the high momentum charged lepton ($\ell = e,\mu$)
from the \Bz\ decay and the slow  pion (\pstar) from the 
\mbox{$\dsm \ra \Dzb \pi^-_s$} decay. 
To reject leptons from semileptonic charm decay and misidentified hadrons, 
the momentum of the lepton candidate in the \FourS\ rest frame ($p^*_\ell$) is required to be in
the range \mbox{$ 1.4 < p^*_\ell < 2.3 ~\gevc$}; 
that of the  \pstar\ ($p^*_\pstar$) has to be less than 0.19 \gevc .
The kinematics of the decay are exploited for further background suppression as follows.
 As a consequence of the limited
phase space available in the decay \mbox{$\dsm \ra \Dzb \pi_s^-$}, the \pstar\ is emitted
within a one-radian wide cone centered about the \dsm\ direction in the \FourS\ rest frame.
The \dsm\ four-momentum can therefore be computed by  approximating its direction as that of
the \pstar , and parameterizing its momentum as a linear function of the \pstar\
momentum, with parameters obtained from the simulation. 
The neutrino invariant mass can be computed from the four-momenta of the \Bz, \dsm, 
and $\ell$ with the relation
\begin{equation}
\nonumber \mnusq = ( P_{\Bz} -  P_{\dsm} -  P_{\ell})^2.
\end{equation}
The momentum of the \Bz\ in the \FourS\ rest frame, on average 0.34 \gevc, is neglected.
\mnusq\ peaks approximately at zero for signal events, whereas background events are spread 
over a wide range. \par
The \Bz\ decay point is determined from a vertex fit of the \pstar\ and $\ell$ tracks,
constrained to  the beam spot position in the plane perpendicular to the beam axis (the $x$-$y$ plane). 
The beam spot is determined on a run-by-run basis using two-prong
events~\cite{ref:babar}. Its size in the horizontal direction is 120 $\mu$m.
Although the beam spot size in the vertical ($y$) direction is only 5.6 $\mu$m, a beam spot 
constraint of 50~$\mu$m is applied to account for the flight of the \Bz\ in the $x$-$y$ plane.
Only events for which the $\chi^2$ probability of the vertex fit, ${\cal P}_V$, is greater than 0.1\% 
are retained. \par
\begin{figure}[!htb]
\begin{center}
\hs{-0.5cm} \includegraphics[height=8cm,width=8.5cm]{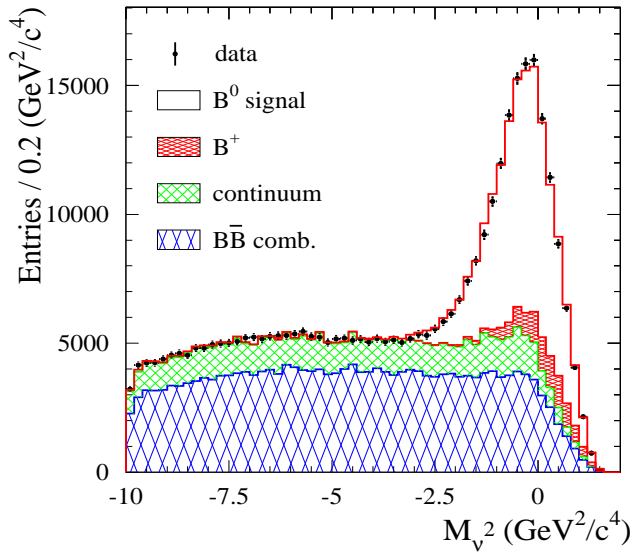} 
\caption{The \mnusq\ spectrum of the selected events. 
The data are represented by the dots with error bars. The result 
of the fit with shapes from the simulation are overlaid. }
\label{fig:mm2_sig}
\end{center}
\end{figure}

 
A selection is applied on the combined likelihood for $p^*_\ell,~p^*_\pstar$, and ${\cal P}_V$, which
results in a signal-to-background  ratio of about one to one in the signal \mnusq\ region, 
defined as \mbox{$\mnusq\ > -2 ~\mathrm{GeV}^2/c^4$}. 
Figure \ref{fig:mm2_sig} shows the \mnusq\ distribution of data events 
used to measure \tBz when the $\ell$ and the \pstar\ have opposite-sign charges. Same-sign
events are used as a background control sample.
The individual distributions shown in Fig.~\ref{fig:mm2_sig} are obtained by fitting to the data the 
contributions from continuum events, obtained from the off-peak data, and from \BB\
combinatorial background,  \Bz\ signal, and \Bu\ resonant background, as predicted by the simulation.
The \Bu\ resonant background is due to intermediate production of higher mass charm resonances
(denoted as \dstrstr ). The fit determines the composition of the selected sample, which
is reported in Table~\ref{t:compo} for the events in the signal region.
\begin{table}
\caption{ Composition of the data sample in the signal region. 
The error is the sum in  quadrature of the statistical
and systematic errors from the fit to the data \mnusq\ distribution. }
\begin{center}
\begin{tabular}{lcc} \hline\hline
Sample     & \# of events & Fraction (\%) \\ \hline
Signal Region  & 172,700 &  -            \\
Backgrounds    &         &               \\
~~~Continuum     & $~19,600 \pm 400  $ & $ 11.4 \pm 0.2$ \\
~~~\BB\ comb.     & $~52,700 \pm 1,400$ & $ 30.0 \pm 0.8$ \\
~~~\Bu        & $~~8,700 \pm 4,400$ & $ ~5.0 \pm 2.5$ \\ 
\hline
\Bz signal & $~91,700 \pm 4,600$ & $ 53.6 \pm 2.6$ \\ \hline\hline
\end{tabular}
\label{t:compo} \end{center} \end{table}
\par The \pep2\ collider produces $\BB$ pairs moving along the beam axis ($z$ direction)
with an average Lorentz boost of $\langle \beta \gamma \rangle = 0.55$. Hence, the two $B$ decay
vertices are separated on average by $\langle \Delta z \rangle \approx 255 ~\mu$m. 
The position of the
\BtoDm\ (``decay'') vertex is reconstructed as described above. The decay point of the
other $B$ is determined from a selection of the remaining tracks in the 
event using the following criteria.
In events that have another lepton with momentum $p^* > 1.1 \gevc$, the other $B$ vertex is computed 
with only this lepton track constrained to the beam-spot in the $x$-$y$ plane.
Otherwise, all the tracks with a \mbox{center-of-mass} angle greater 
than 90$^\circ$ with respect to the \pstar\ direction are
considered. This requirement is used to remove most of the tracks from the decay of  the \Dzb\ daughter 
of the \dsm, which would otherwise bias the reconstruction of the other $B$ vertex position.   
Simulation shows that  in about 75\% of signal events the other  
vertex has no tracks from the \Dzb\ decay.
The selected tracks are then constrained to  the beam-spot in the $x$-$y$ plane.
The track with the largest contribution to the vertex $\chi^2$, if greater than 6, is removed 
and the fit iterated until no track fails this requirement.
Vertices composed of just one track that is not a high momentum lepton
 are rejected in order to reduce the number of poorly measured vertices.
The lifetime is determined by measuring the quantity 
$\Delta z = z_{\mathrm{decay}} - z_{\mathrm{other}}$, where $z_{\mathrm{decay}} (z_{\mathrm{other}})$
is the position along the beam line of the decay (other) vertex. 
The proper time difference is then computed 
with the relation \mbox{$\Delta t = \Delta z / (c \beta\gamma )$}. A fit with a double
Gaussian to the $\Delta t$ residuals in the Monte Carlo simulation shows that one half of the events are 
contained in the narrower Gaussian, which has a width of 0.7~\ps . 
To remove badly reconstructed vertices, all events
for which either $|\Delta z| > 3$ mm or $\sigma_{\Delta z}> 500 ~\mu$m are rejected, where
$\sigma_{\Delta z}$ is the uncertainty on $\Delta z $ computed
for each event. \par
\tBz is obtained from a binned maximum likelihood fit to the two-dimensional $\Delta t,\sigma_{\Delta t}$ distribution.
To save computation time events are grouped in a two-dimensional space consisting of 100 $\Delta t$ 
and 25 $\sigma_{\Delta t}$ bins.
The $\Delta t$ distribution of signal events,  ${\cal F}(\Delta t,\sigma_{\Delta t},\tBz)$, 
is described by the convolution of the decay probability distribution
\begin{eqnarray}
\nonumber f(\Delta t_{\rm true} | \tBz) = \frac{1}{2\tBz}  exp(-|\Delta t_{\rm true}|/\tBz),
\end{eqnarray}
with the experimental resolution function, which is parametrized by the sum of three
Gaussian distributions. The two narrow Gaussians, which account for more than 99\% of the events,
have the form
\begin{eqnarray}
\nonumber {\cal G}(\delta(\Delta t), \sigma_{\Delta t}) \equiv 
\frac{1}{\sqrt{2\pi}S\sigma_{\Delta t}} 
exp(- \frac {(\delta(\Delta t) -b)^2}{2{S^2 \sigma^2_{\Delta t}}}) ~,
\end{eqnarray}
where \mbox{$\delta (\Delta t) = \Delta t - \Delta t_{\rm true}$} is the difference between the 
measured and the true value of $\Delta t$, $b$ is a bias due to the charm tracks in the other 
vertex and resolution effects,  and the scale factor $S$ is introduced to account for possible 
misestimation of the calculated error $\sigma_{\Delta t}$ on the proper time difference. 
The third Gaussian of fixed bias ($-2 \ps$) and width ($8 \ps$) is added to account for badly 
measured events (``outliers'').\par
\Bu\ background events that peak in the \mnusq\ signal region are described by an identical function, 
with the same resolution parameters as for the \Bz\ signal events, and an effective \Bu\ lifetime
of 1.57 \ps.  This value,  obtained by fitting simulated \Bu\ events, 
is smaller than the value of 1.655 \ps generated in the simulation
due to \Dzb tracks from the decay vertex being included in the other vertex. \par
The $\Delta t$ distribution of continuum background events is modeled as the sum of two 
components, one with non-zero lifetime and the other with zero lifetime, 
convolved with the same single Gaussian resolution function.
The parameters of the resolution function, as well as the lifetime and the fraction
of events with non-zero lifetime, are all determined with the off-peak events that satisfy the
selection criteria. \par
The $\Delta t$ distribution of the combinatorial $\BB$ background is modeled as the sum of
a non-zero and a zero-lifetime component, with a resolution function that is the sum of 
three Gaussians. All parameters are determined from the data by 
fitting the measured \deltat\ distribution of the events in the
sideband region, defined by $-10 < \mnusq < -4 ~\mathrm{GeV}^2/c^4$. 
The Monte Carlo simulation
shows, however, that there are small differences in the lifetime and in the fraction of 
events with non-zero lifetime between the signal region and the sideband.
These differences are also observed in the data 
by separately fitting signal region and sideband events in the same sign $\ell \pstar$ background 
control sample. The results from the like-sign fits are used to scale the two background 
parameters from the sideband to the signal region.\par
The function used to fit the data is the weighted sum of the four
contributions:
\begin{widetext}
\begin{eqnarray*}
{\cal F}(\Delta t,\sigma_{\Delta t}|\tBz) 
& = & [1 - f_{\Bu} (\mnusq) - f_{c}(\mnusq) - f_{\BB}(\mnusq) ]
\,{\cal F}_{\Bz} (\Delta t,\sigma_{\Delta t},\tBz) + 
f_{\Bu} (\mnusq)\, {\cal F}_{\Bu} (\Delta t,\sigma_{\Delta t}) + \\
&& f_{c}(\mnusq)\, {\cal F}_{c}~(\Delta t,\sigma_{\Delta t}) +
f_{\BB}(\mnusq)\, {\cal F}_{\BB}(\Delta t,\sigma_{\Delta t}), 
\end{eqnarray*}
\end{widetext}
where the functions ${\cal F}_{\Bz},~{\cal F}_{\Bu},~{\cal F}_{c}$~ and ${\cal F}_{\BB}$ 
describe the measured decay time difference distributions
for the signal, peaking \Bu,  continuum, and $\BB$ combinatorial background, 
respectively. $f_{\Bu},~f_{c}$ and $f_{\BB}$ are the probabilities that the event 
is from the \Bu, continuum, or $\BB$ background, computed for each event on the basis of the
measured value of \mnusq . 
Simultaneously with $\tau_{B^0}$, the following parameters of the signal resolution function
are fitted: the scale factor of the first Gaussian, \mbox{$S_{1} = 1.02 \pm 0.02$}, 
the scale factor of the second Gaussian, \mbox{$S_{2} = 2.4 \pm 0.1$}, the bias of the first 
Gaussian, \mbox{$b_1 = -0.120 \pm 0.009$}, and the fraction of outliers, \mbox{$f_o = (0.2\pm0.1)\%$}. 
The fraction of events contained in the second Gaussian, $f_2$, and 
its bias $b_2$ are fixed to $7\%$ and $-0.85$~\ps, respectively. \par
\begin{figure}[!htb]
\begin{center} 
\hs{-0.5cm}\includegraphics[height=7cm,width=8.5cm]{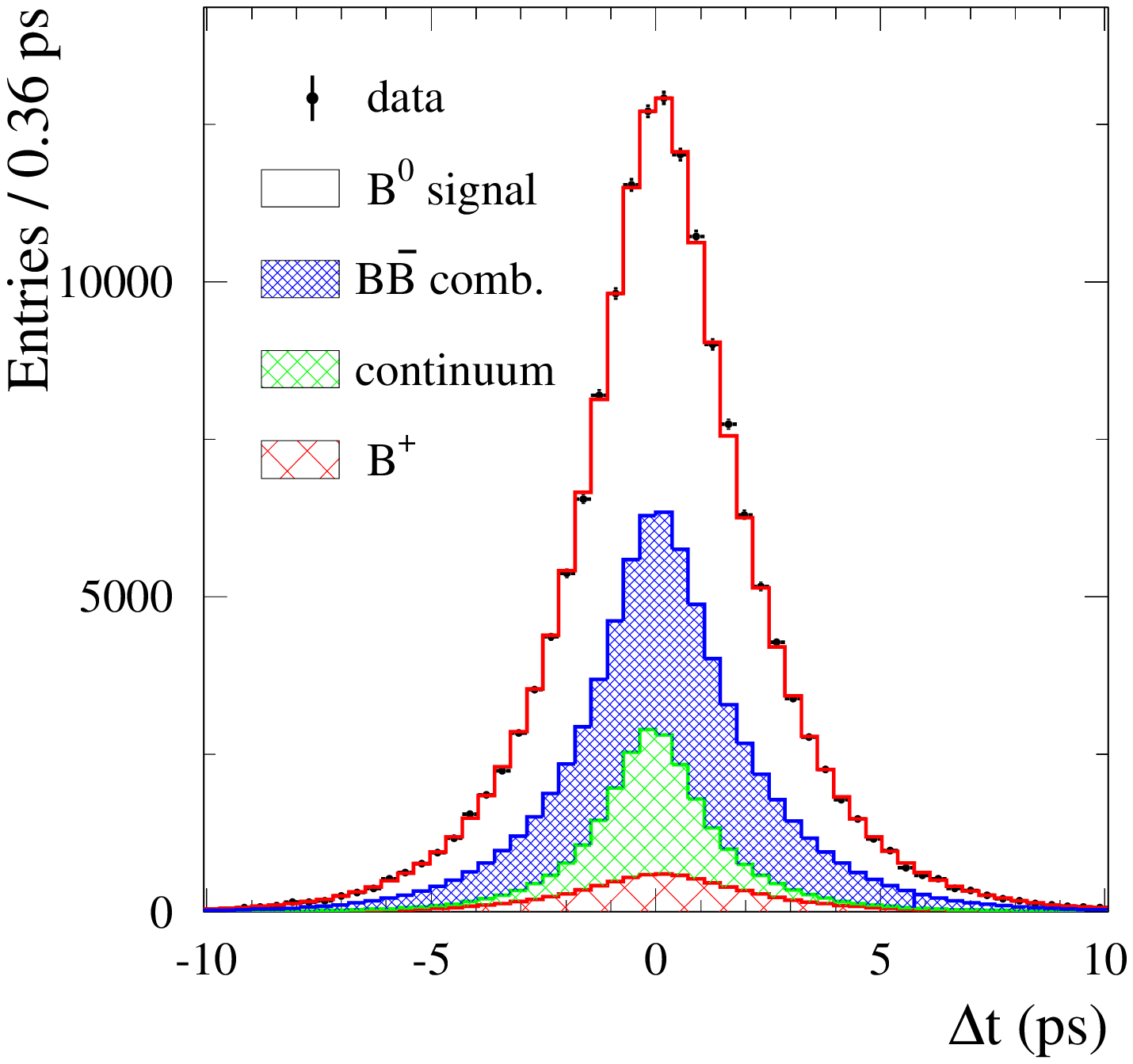} \\
\hs{-0.5cm}\includegraphics[height=7cm,width=8.5cm]{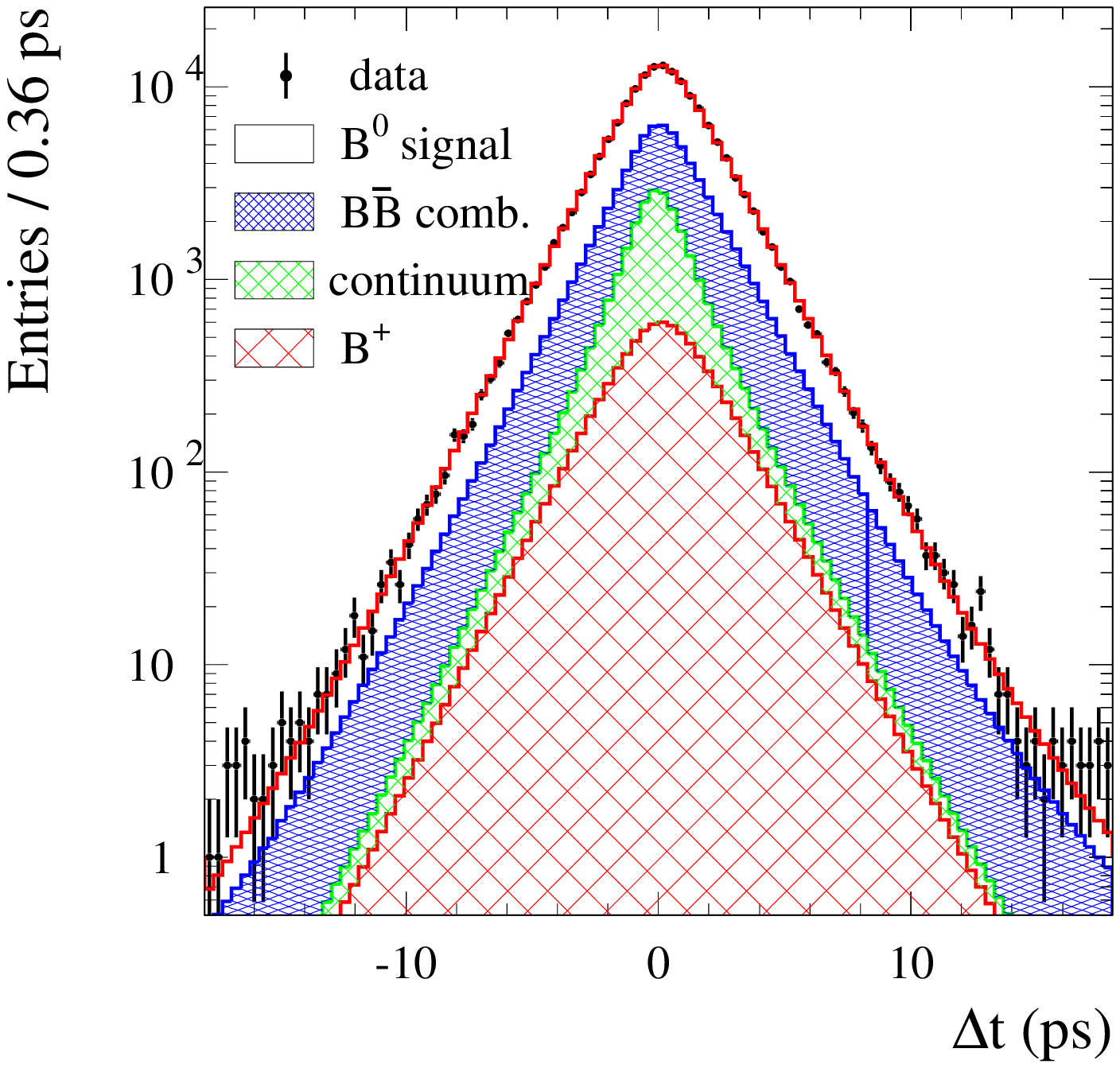} \\
\caption{$\Delta t$ distribution for selected events in the data (points) in linear (upper)
and logarithmic (lower) scale.
The lifetime fit result is superimposed on the data.  
The hatched histograms show the contributions from the background sources described in the text.
}
\label{f:tau}
\end{center}
\end{figure}

The result of the fit is
\mbox {$ \tau_{\Bz}^{\mathrm{raw}} = 1.482 \pm 0.012 \ps$},
where the error is statistical only. Figure \ref{f:tau} shows the comparison between the
measured $\Delta t$ distribution and the fit result.
The probability of obtaining a lower likelihood, evaluated with a Monte Carlo technique, is 18\%.
This raw lifetime must be corrected for the bias induced by the tracks from the 
\Dzb\ that are not rejected by the \pstar\ cone cut. A multiplicative correction factor of 
\mbox{${\cal R}_{\Dzb} = 1.032 \pm 0.007 \mathrm{(stat)} \pm 0.007 \mathrm{(syst)}$} is computed from the
simulation. The statistical error arises from the number of simulated events.
The dominant systematic uncertainty corresponds to the full variation in ${\cal R}_{\Dzb}$ (0.66\%)
 obtained by smearing the \deltat\ resolution in the simulation to match that in the data. 
A second systematic uncertainty is computed
 by comparing in data and simulation  the fraction of charged tracks
from \Dzb\ decays outside the \pstar\ cone for a subset of events in which the \Dzb\ is fully 
reconstructed in the $K^+\pi^-$,
$K^+\pi^-\pi^0$, and $K^+\pi^-\pi^+\pi^-$ final states. 
The maximum discrepancy between data and simulation
corresponds to a variation of $\pm 0.24 \%$ in the value of
${\cal R}_{\Dzb}$.
The corrected value of the \Bz\ lifetime is then
\ba
\nonumber \tBz = \tau_{\Bz}^{\mathrm{raw}} {\cal R}_{\Dzb} = 1.529 \pm 0.012 \ps.
\ea
The systematic error on \tBz is computed by adding in quadrature the contributions from several sources,
described below and summarized in Table \ref{t:syste}.\par
The fractions of \Bu, continuum, and combinatorial $\BB$ events are varied by the uncertainties obtained from the
\mnusq\ fit (see Table \ref{t:compo}). The parameters of the continuum and combinatorial $\BB$ \deltat
distributions are varied by their uncertainties, accounting for their
correlations. As described above, the fraction of events with non-zero lifetime and the lifetime of the combinatorial 
$\BB$ background computed from the sideband are corrected 
with the same-sign control sample. This method is validated by a simulation study, and the statistical
error of the validation is included in the background systematic error. The effective \Bu\ lifetime is
varied by $\pm 3\%$, which is the sum in quadrature of the world average error on the \Bu\ 
lifetime and the statistical and systematic uncertainties of the \Dzb\ bias correction. \par
The parameters of the signal resolution function that are not determined in the fit to the data are varied 
within conservative ranges ($f_2$ between 0.03 and 0.13, and $b_2$ between $-1.5$ and 0 \ps). 
Several different analytical expressions are used to represent the small fraction of outliers.
The fit is also performed by allowing the scale factor and the bias of the narrow Gaussians to depend 
linearly on $\sigma_{\deltat}$ or on the lepton polar angle.
The maximum change with respect to the result with fixed parameters is taken as the systematic error
due to the parametrization of the resolution function. 
\par
The bias due to the event selection is found to be compatible with zero by fitting with an exponential function
the true proper time difference of signal events selected in the simulation.
The statistical error of this test is added to the systematic error.\par
The statistical and systematic errors on ${\cal R}_{\Dzb}$ are propagated to the final error. A possible bias 
induced by the presence of tracks from charm decays produced by the other $B$ meson is investigated in
the simulation by varying within their uncertainties the relative fractions of charmless, single charm, and 
double charm events, and also by varying the relative fractions of $D^+,D^0,D_s$, and $\Lambda_c$ hadrons.
The $z$ length scale is determined with an uncertainty of 0.4\% from secondary interactions with 
a beam pipe section of known length.
The dependence of the result on several different variables 
(angular width of the \pstar\ cone used to reject \Dzb tracks, 
soft pion momentum, lepton momentum, polar and azimuthal angle, 
alignment conditions) is carefully inspected; no statistically significant effect is found.
No difference in the result is observed if \tBz is determined with an unbinned maximum likelihood fit.
A final relative error of $\pm 1.9\%$ is found by adding in quadrature the uncertainties from the above sources,
as listed in Table \ref{t:syste}.\par
In conclusion, a sample of about 92000 \BtoDm\ decays is selected by partial reconstruction
of the \mbox{$\dsm \rightarrow \Dzb \pi^- $ } decay. It is used for
a measurement of the \Bz\ lifetime. The value obtained,
\ba
\nonumber \tBz = 1.529 \pm 0.012 ~\mathrm{(stat)} \pm 0.029 ~\mathrm{(syst)} \ps,
\ea
is consistent with a recent \babar\ measurement \cite{tauBabar} and with the world average 
\cite{PDG}. It is currently the most precise single measurement of this quantity.

\begin{table}[hb]
\caption{Contributions to the systematic error.}
\begin{center}
\begin{tabular}{lc} \hline \hline 
Source & $\sigma_{\tBz}/{\tBz}(\%)$ \\ \hline 
Continuum fraction $\&$ parametrization & 0.36 \\ 
$\BB$ fraction $\&$ parametrization& 0.68 \\ 
\Bu fraction $\&$ parametrization  & 0.64 \\ 
Resolution model & 1.14 \\  
Event selection bias & 0.30 \\
\Dzb\ bias (${\cal R}_{\Dzb}$)  	 & 0.95 \\
Bias due to charm from the other $B$ & 0.21 \\ 
$z$ scale          & 0.40 \\  \hline
Total 		 & 1.89 \\   \hline \hline 
\end{tabular} \label{t:syste}\end{center} \end{table}
We are grateful for the excellent luminosity and machine conditions
provided by our \pep2\ colleagues.
The collaborating institutions wish to thank 
SLAC for its support and kind hospitality. 
This work is supported by
DOE
and NSF (USA),
NSERC (Canada),
IHEP (China),
CEA and
CNRS-IN2P3
(France),
BMBF
(Germany),
INFN (Italy),
NFR (Norway),
MIST (Russia), and
PPARC (United Kingdom). 
Individuals have received support from the Swiss NSF, 
A.~P.~Sloan Foundation, 
Research Corporation,
and Alexander von Humboldt Foundation.

\end{document}